\documentclass[12pt,preprint,preprintnumbers,nofootinbib]{revtex4}
\usepackage{graphicx}
\usepackage{epsfig}    
\usepackage{dcolumn}
\usepackage{bm}
\usepackage{amssymb}
\usepackage{epsfig}    
\usepackage{here}
\begin{document}
\topmargin -1cm

\bibliographystyle{revtex}

\title{Update analysis of two-body charmed $B$ meson decays}

\author{Cheng-Wei Chiang$^{1,2}$\footnote{E-mail: chengwei@phy.ncu.edu.tw} and
  Eibun Senaha$^1$\footnote{E-mail: senaha@ncu.edu.tw}}
\affiliation{$^1$Department of Physics, National Central University, Chungli,
  Taiwan 320, R.O.C}
\affiliation{$^2$Institute of Physics, Academia Sinica, Taipei, Taiwan 115,
  R.O.C.}  \bigskip

\date{\today}

\begin{abstract}
  The charmed $B$ decays, $B\to DP, ~D^*P$ and $DV$, are re-analyzed using the
  latest experimental data, where $P$ and $V$ denote the pseudoscalar meson and
  vector meson, respectively.  We perform global fits under the assumption of
  flavor SU(3) symmetry.  The size of the decay amplitudes and the strong
  phases between the topologically distinct amplitudes are studied. Predictions
  of the related $B_s$ decay rates are made based upon the fitted results.  We
  also note a serious SU(3) symmetry breaking or inconsistency in the $DV$
  sector.
\end{abstract}
\maketitle

\section{Introduction \label{sec:intro}}

The hadronic decays of $B$ mesons have provided us with a good place to study
CP violation in particle physics.  In particular, the detection of direct CP
violation in a decay process requires that there exist at least two
contributing amplitudes with different weak and strong phases.  The direct
CP-violating effect in the $B$ system has finally been observed in the $B^0 \to
K^+ \pi^-$ decay at the $B$-factories \cite{Aubert:2004qm,Chao:2004mn}, proving
the existence of nontrivial strong phases in $B$ decays.  It is therefore of
consequence to find out the patterns of final-state strong phases for a wider
set of decay modes.

Since the CKM factors involved in charmed $B$ meson decays are purely real to a
good approximation, the phases associated with the decay amplitudes thus have
the origin of strong interactions.  Such final-state rescattering effects have
been noticed from data in these decays \cite{Cheng:2001sc,Chua:2001br}, and
estimated to be at 15-20\% level \cite{Fayyazuddin:2002pv}.  Unfortunately, no
satisfactory first-principle calculations can yield such strong phases
\cite{Wolfenstein:2003pc}.  In Ref.~\cite{Chiang:2002tv}, we performed an
analysis based upon the experimental data available at that time.  A few
theoretical and experimental questions are left unanswered.  As more decay
modes have been observed and others are measured at higher precisions, it
becomes possible for us to look at and answer those questions.  In this paper,
flavor SU(3) symmetry is employed to relate different amplitudes and strong
phases of the same topological type.  Moreover, we will take a different
approach by fitting theoretical parameters to all available branching ratios
simultaneously.  An advantage of this analysis is that the parameters thus
obtained are insensitive to statistical fluctuations of individual modes.

This paper is organized as follows.  In Section~\ref{sec:decomp}, we give the
amplitude decomposition of modes under flavor SU(3) symmetry and the current
branching ratio data.  Theoretical parameters involved in our analysis are
defined.  In Section~\ref{sec:phases}, we consider three sets of charmed decay
modes: $DP$, $D^* P$, and $DV$, where $P$ and $V$ denote charmless pseudoscalar
and vector mesons, respectively.  A summary of our findings is given in
Section~\ref{sec:summary}.

\section{Flavor Amplitude Decomposition and Data \label{sec:decomp}}

In the decomposition of decay amplitudes, relevant meson wave functions are
assumed to have the following quark contents, with phases chosen so that
isospin multiplets contain no relative signs:

\begin{itemize}
\item{\it Beauty mesons:} $\overline{B^0} = b \bar d$, $B^- = - b \bar u$,
$\overline{B}_s = b \bar s$.
\item {\it Charmed mesons:} $D^0 = - c \bar u$, $D^+ = c \bar d$, $D_s^+ =c
  \bar s$, with corresponding phases for vector mesons.
\item {\it Pseudoscalar mesons $P$:} $\pi^+ = u \bar d$, $\pi^0 = (d \bar d - u
  \bar u)/\sqrt{2}$, $\pi^- = - d \bar u$, $K^+ = u \bar s$, $K^0 = d \bar s$,
  $\bar K^0 = s \bar d$, $K^- = - s \bar u$, $\eta = (s \bar s - u \bar u - d
  \bar d)/\sqrt{3}$, $\eta' = (u \bar u + d \bar d + 2 s \bar s)/\sqrt{6}$,
  assuming a specific octet-singlet mixing \cite{Chau:1990ay,eta} in the $\eta$
  and $\eta'$ wave functions.)
\item {\it Vector mesons $V$:} $\rho^+ = u \bar d$, $\rho^0 = (d \bar d - u
  \bar u)/\sqrt{2}$, $\rho^- = - d \bar u$, $\omega = (u \bar u + d \bar
  d)/\sqrt{2}$, $K^{*+} = u \bar s$, $K^{*0} = d \bar s$, $\overline{K}^{*0} =
  s \bar d$, $K^{*-} = - s \bar u$, $\phi = s \bar s$.
\end{itemize}

The amplitudes contributing to the decays discussed here involve only three
different topologies \cite{Zeppenfeld:1980ex,Chau:1990ay,GHLR,eta}:
\begin{enumerate}
\item {\it Tree amplitude $T$:} This is associated with the transition $b \to c
  d \bar u$ (Cabibbo-favored) or $b \to c s \bar u$ (Cabibbo-suppressed) in
  which the light (color-singlet) quark-antiquark pair is incorporated into one
  meson, while the charmed quark combines with the spectator antiquark to form
  the other meson.
\item {\it Color-suppressed amplitude $C$:} The transition is the same as in
  the tree amplitudes, namely $b \to c d \bar u$ or $b \to c s \bar u$, except
  that the charmed quark and the $\bar u$ combine into one meson while the $d$
  or $s$ quark and the spectator antiquark combine into the other meson.
\item{\it Exchange amplitude $E$:} The $b$ quark and spectator antiquark
  exchange a $W$ boson to become a $c \bar u$ pair, which then hadronizes into
  two mesons by picking up a light quark-antiquark pair out of the vacuum.
\end{enumerate}

After factoring out the CKM factors explicitly, we obtain the flavor amplitude
decomposition of the charmed $B$ decay modes in Tables~\ref{tab:DP},
\ref{tab:DstP}, and \ref{tab:DV}. In these tables, we introduce positive
$\xi$'s to parameterize the flavor SU(3) breaking effects. This symmetry is
respected between strangeness-conserving and strangeness-changing amplitudes
when $\xi$'s are taken to be unity.  As we will discuss in the next section,
$\xi$'s will be allowed to change in order to test the assumption.  Using the
Wolfenstein parameters \cite{Wolfenstein:1983yz}, the relevant CKM factors are:
\begin{eqnarray}
  V_{cb} = A \lambda^2 ~,
  \quad
  V_{ud} = 1 - \frac{\lambda^2}{2} ~,
  \quad \mbox{and} \quad
  V_{us} = \lambda ~,
\end{eqnarray}
none of which contain a weak phase to the order we are concerned with.  In the
following analysis, we take the central values $\lambda = 0.2272$ and $A =
0.809$ quoted by the CKMfitter group \cite{CKMfitter}.

Since only the relative strong phases are physically measurable, we fix the
tree ($T$, $T_P$, and $T_V$) amplitudes to be real and pointing in the positive
direction.  We then associate the color-suppressed and exchange amplitudes with
the corresponding strong phases explicitly as follows:
\begin{eqnarray}
  C = |C| e^{i\delta_{C}} ~, && E = |E| e^{i\delta_{E}} ~, \\
  C_P = |C_P| e^{i\delta_{C_P}} ~, && E_P = |E_P| e^{i\delta_{E_P}} ~, \\
  C_V = |C_V| e^{i\delta_{C_V}} ~, && E_V = |E_V| e^{i\delta_{E_V}} ~.
\end{eqnarray}

The magnitude of invariant decay amplitude ${\cal A}$ for a decay process $B
\to M_1 \, M_2$ is related to its partial width via the following relation:
\begin{eqnarray}
  \Gamma(B \to M_1 \, M_2)
  = \frac{p^*}{8\pi m_B^2} |{\cal A}|^2 ~,
\end{eqnarray}
with
\begin{equation}
p^* = \frac{1}{2m_B}
\sqrt{\big\{m_B^2-(m_1+m_2)^2\big\}\big\{m_B^2-(m_1-m_2)^2\big\} }~,
\end{equation}
where $m_{1, 2}$ are the masses of $M_{1, 2}$, respectively.  To relate partial
widths to branching ratios, we use the world-average lifetimes $\tau^+ = (1.638
\pm 0.011)$ ps, $\tau^0 = (1.530 \pm 0.009)$ ps, and $\tau_s = (1.466 \pm
0.059)$ ps computed by the Heavy Flavor Averaging Group (HFAG) \cite{HFAG}.

\begin{table}[t]
\caption{Branching ratios and flavor amplitude decomposition for $B\to DP$
  decays.  Data are quoted from Refs. \cite{PDG, Aubert:2006hu,
  Ronga:2006hv, Aubert:2006qn, Aubert:2006um, Blyth:2006at, Aubert:2006jc,
  Aubert:2006cd, Kuzmin:2006mw,unknown:2006qw}}
\label{tab:DP}
\begin{center}
\begin{tabular}{l c c c c c} \hline \hline
Decay & $m_B$   & Branching ratio & $p^*$ & $|{\cal A}|$ & Representation \\
      & (GeV) & (in units of $10^{-4}$) & (GeV) & ($10^{-7}$ GeV) \\
\hline
$B^- \to D^0 \pi^-$ & 5.2791 & $47.5 \pm 2.1 $ & 2.308 & $ 7.61 \pm 0.17 $
                        & $-V_{cb}V_{ud}^*(T+C)$ \\
$    \to D^0 K^-$   &        & $ 4.08 \pm 0.24 $ & 2.281 & $ 2.24 \pm 0.07 $
                        & $-V_{cb}V_{us}^* (\xi_TT+\xi_CC)$ \\
\hline
$\overline{B}^0 \to D^+ \pi^-$ & 5.2793 & $ 29 \pm 2 $ & 2.306 & $ 6.11 \pm 0.21 $
                        & $-V_{cb}V_{ud}^*(T+E) $ \\
$   \to  D^+ K^-$   &        & $2.0 \pm 0.6$ & 2.279 & $1.63 \pm 0.24 $
                        & $-V_{cb}V_{us}^* \xi_T T$ \\
$    \to D^0 \pi^0$ &        & $ 2.61 \pm 0.24 $ & 2.308 & $1.85 \pm 0.09 $
                        & $V_{cb}V_{ud}^*(E-C)/\sqrt{2}$ \\
$   \to  D^0 \eta$  &   & $2.0\pm0.2$ & 2.274 & $1.62 \pm 0.08 $
                        & $V_{cb}V_{ud}^*(C+E)/\sqrt{3}$ \\
$   \to  D^0 \eta'$ &        &     $ 1.25 \pm 0.23 $  & 2.198 & $1.31 \pm 0.12 $
                        & $-V_{cb}V_{ud}^*(C+E)/\sqrt{6}$ \\
$   \to D^0 \overline{K}^0$ & & $0.52 \pm 0.07$ & 2.280 & $0.83 \pm 0.05 $
                        & $-V_{cb}V_{us}^* \xi_C C$ \\
$   \to  D_s^+ K^-$ & & $0.27 \pm 0.05$ & 2.242 & $0.61 \pm 0.06 $
                        & $-V_{cb}V_{ud}^*E$ \\
                        \hline
$\overline{B}^0_s \to D^+ \pi^-$ & 5.3696 & $   $ & 2.357  & $  $
                        & $-V_{cb}V_{us}^*\xi_E E $ \\ 
$    \to D^0 \pi^0$ &        & $  $ & 2.359 & $ $
                        & $V_{cb}V_{us}^*\xi_E E/\sqrt{2}$ \\
$    \to D^0 K^0$ &        & $  $ & 2.332 & $ $
                        & $-V_{cb}V_{ud}^*C$\\
$   \to  D^0 \eta$  &   & $ $ & 2.326  & $$
                        & $V_{cb}V_{us}^*(\xi_E E-\xi_CC)/\sqrt{3}$ \\
$   \to  D^0 \eta'$ &        &     $  $  & 2.251 & $ $
                        & $-V_{cb}V_{us}^*(2\xi_CC+\xi_EE)/\sqrt{6}$ \\
$   \to  D_s^+ \pi^-$ & & $38\pm3\pm13^{~a} $ & 2.321 & $ 7.30 \pm 1.28 $
                        & $-V_{cb}V_{ud}^*T$ \\
$   \to  D_s^+ K^-$ & & $  $ & 2.294 & $ $
                        & $-V_{cb}V_{us}^*(\xi_TT+\xi_EE)$ \\
\hline\hline                      
\end{tabular}
\end{center}
\leftline{$^a$ Ref.~\cite{unknown:2006qw}.}
\end{table}

\begin{table}[t]
\caption{Branching ratios and flavor amplitude decomposition for $B\to D^*P$
  decays.  Data are quoted from Refs. \cite{PDG, Aubert:2006hu,
  Ronga:2006hv, Aubert:2006qn, Aubert:2006um, Blyth:2006at, Aubert:2006jc,
  Aubert:2006cd, Kuzmin:2006mw,unknown:2006qw}}
\label{tab:DstP}
\begin{center}
\begin{tabular}{l c c c c c} \hline \hline
Decay & $m_B$   & Branching ratio & $p^*$ & $|{\cal A}|$ & Representation \\
      & (GeV) & (in units of $10^{-4}$) & (GeV) & ($10^{-7}$ GeV) \\
\hline
$B^- \to D^{*0} \pi^-$ & 5.2791 & $ 50 \pm 4 $ & 2.256 & $7.87 \pm 0.32 $
                        & $-V_{cb}V_{ud}^*(T_V+C_P)$ \\
$    \to D^{*0} K^-$   &        & $ 3.7 \pm 0.4 $ & 2.227 & $2.16 \pm 0.12 $
                        & $-V_{cb}V_{us}^* (\xi_{T_V}T_V+\xi_{C_P}C_P)$ \\
\hline
$\overline{B}^0 \to D^{*+} \pi^-$ & 5.2793 & $28.5 \pm 1.7$ & 2.255 & $ 6.17 \pm 0.19 $
                        & $-V_{cb}V_{ud}^*(T_V+E_P) $ \\
$   \to  D^{*+} K^-$   &        & $ 2.14 \pm 0.20 $ & 2.226 & $1.70 \pm 0.08 $
                        & $-V_{cb}V_{us}^* \xi_{T_V} T_V$ \\
$    \to D^{*0} \pi^0$ &        & $1.7 \pm 0.3 $ & 2.256 & $1.52 \pm 0.12 $
                        & $V_{cb}V_{ud}^*(E_P-C_P)/\sqrt{2}$ \\
$   \to  D^{*0} \eta$  &   & $ 1.8 \pm 0.6 $ & 2.220 & $ 1.55 \pm 0.24 $
                        & $V_{cb}V_{ud}^*(C_P+E_P)/\sqrt{3}$ \\
$   \to  D^{*0} \eta'$ &        &     $ 1.23 \pm 0.35 $  & 2.141 & $ 1.32 \pm 0.19 $
                        & $-V_{cb}V_{ud}^*(C_P+E_P)/\sqrt{6}$ \\
$   \to D^{*0} \overline{K}^0$    & & $ 0.36 \pm 0.12^{~b} $ & 2.227 & $ 0.70 \pm 0.12 $
                        & $-V_{cb}V_{us}^* \xi_{C_P} C_P$ \\
$   \to  D_s^{*+} K^-$ & & $  0.20 \pm 0.05 \pm 0.04^{~c} $ & 2.185 & $ 0.53 \pm 0.08 $
                        & $-V_{cb}V_{ud}^*E_P$ \\
\hline
$\overline{B}^0_s \to D^{*+} \pi^-$ & 5.3696 & $ $ & 2.306  & $  $
                        & $-V_{cb}V_{us}^*\xi_{E_P}E_P $ \\ 
$    \to D^{*0} \pi^0$ &        & $ $ & 2.308 & $ $
                        & $ V_{cb}V_{us}^*\xi_{E_P} E_P/\sqrt{2}$ \\
$    \to D^{*0} K^0$ &        &  & 2.279 & $ $
                        & $-V_{cb}V_{ud}^*C_P$ \\
$   \to  D^{*0} \eta$  &   & & 2.273 & $$
                        & $V_{cb}V_{us}^*(\xi_{E_P}E_P-\xi_{C_P}C_P)/\sqrt{3}$ \\
$   \to  D^{*0} \eta'$ &        &     $ $  & 2.195 & $ $
                        & $-V_{cb}V_{us}^*(2\xi_{C_P}C_P+\xi_{E_P}E_P)/\sqrt{6}$ \\
$   \to  D_s^{*+} \pi^-$ & & $  $ & 2.267 & $ $
                        & $-V_{cb}V_{ud}^*T_V$ \\
$   \to  D_s^{*+} K^-$ & & $  $ & 2.238 & $ $
                        & $-V_{cb}V_{us}^*(\xi_{T_V}T_V+\xi_{E_P}E_P)$ \\
\hline\hline                      
\end{tabular}
\end{center}
\leftline{$^b$ Ref.~\cite{Aubert:2006qn}, $^c$ Ref.~\cite{Aubert:2006hu}.}
\end{table}

\begin{table}
\caption{Branching ratios and flavor amplitude decomposition for $B\to DV$
  decays.  Data are quoted from Refs. \cite{PDG, Aubert:2006hu,
  Ronga:2006hv, Aubert:2006qn, Aubert:2006um, Blyth:2006at, Aubert:2006jc,
  Aubert:2006cd, Kuzmin:2006mw,unknown:2006qw}}
\label{tab:DV}
\begin{center}
\begin{tabular}{l c c c c c} \hline \hline
Decay & $m_B$   & Branching ratio & $p^*$ & $|{\cal A}|$ & Representation \\
      & (GeV) & (in units of $10^{-4}$) & (GeV) & ($10^{-7}$ GeV) \\
\hline
$B^- \to D^0 \rho^-$ & 5.2791 & $134 \pm 18 $ & 2.237 & $13.0 \pm 0.9 $
                        & $-V_{cb}V_{ud}^*(T_P+C_V)$ \\
$    \to D^0 K^{*-}$   &        & $ 5.3 \pm 0.4 $ & 2.213 & $2.60 \pm 0.11 $
                        & $-V_{cb}V_{us}^* (\xi_{T_P}T_P+\xi_{C_V}C_V)$ \\
\hline
$\overline{B}^0 \to D^+ \rho^-$ & 5.2793 & $ 75 \pm 12 $ & 2.235 & $10.1 \pm 0.8 $
                        & $-V_{cb}V_{ud}^*(T_P+E_V) $ \\
$   \to  D^+ K^{*-}$   &        & $ 4.5 \pm 0.7 $ & 2.211 & $2.48 \pm 0.19 $
                        & $-V_{cb}V_{us}^* \xi_{T_P} T_P$ \\
$\to D^0 \rho^0$ &      & $3.2 \pm 0.5 $ & 2.237 & $2.07 \pm 0.16 $
                        & $V_{cb}V_{ud}^*(E_V-C_V)/\sqrt{2}$ \\
$   \to  D^0 \omega$  &   & $2.6 \pm 0.3$ & 2.235 & $1.87 \pm 0.11 $
                        & $-V_{cb}V_{ud}^*(C_V+E_V)/\sqrt{2}$ \\
$\to D^0 \overline{K}^{*0}$ & & $ 0.42 \pm 0.06 $ & 2.212 & $0.76 \pm 0.06 $
                        & $-V_{cb}V_{us}^* \xi_{C_V} C_V$ \\
$   \to  D_s^+ K^{*-}$ & & $ < 8 $ & 2.172 & $ < 3 $
                        & $-V_{cb}V_{ud}^*E_V$ \\
\hline
$\overline{B}^0_s \to D^+ \rho^-$ & 5.3696 & $   $ & 2.288 & $  $
                        & $-V_{cb}V_{us}^*\xi_{E_V} E_V $ \\ 
$   \to  D^+ K^{*-}$ & & $  $ & 2.264 & $ $
                        & $-V_{cb}V_{us}^*(\xi_{T_P}T_P+\xi_{E_V}E_V)$ \\
$\to D^0 \rho^0$ &      &  & 2.289 & $ $
                        & $V_{cb}V_{ud}^*E_V/\sqrt{2}$ \\                        
$    \to D^0 K^{*0}$ &        &  & 2.265 & $ $
                        & $-V_{cb}V_{ud}^*C_V$ \\
$   \to  D^0 \omega$  &   &  & 2.288 & $$
                        & $-V_{cb}V_{us}^*\xi_{E_V} E_V/\sqrt{2}$ \\
$   \to  D^0 \phi$  &   & & 2.237 & $$
                        & $-V_{cb}V_{us}^*\xi_{C_V} C_V$ \\
$   \to  D_s^+ \rho^-$ & & & 2.250 & $ $
                        & $-V_{cb}V_{ud}^*T_P$ \\
$   \to  D_s^+ K^{*-}$ & & & 2.226 & $ $
                        & $-V_{cb}V_{us}^*(\xi_{T_P}T_P+\xi_{E_V}E_V)$ \\                        
\hline\hline                      
\end{tabular}
\end{center}
\end{table}

\section{Strong Phases \label{sec:phases}}

In our analysis, we take the amplitude sizes and the strong phases as
theoretical parameters, and perform $\chi^2$ fits to all the branching ratios
in each category ($B_{u, d}\to DP$, $D^*P$, and $DV$). We consider three
schemes to test the flavor SU(3) assumption:
\begin{enumerate}
\item $\xi_T=\xi_C=\xi_{T_V}=\xi_{C_P}=\xi_{T_P}=\xi_{C_V}=1$.  This is the
  exact flavor SU(3)-symmetric case.
\item $\xi_T=\xi_C=\xi_{T_V} = \xi_{C_P}= f_K / f_\pi \simeq 1.22$, and
  $\xi_{T_P}=\xi_{C_V}= f_{K^*} / f_{\rho} \simeq 1.00 $.  This takes into
  account the difference in the decay constants for the charmless meson in the
  final states.
\item All $\xi_T$'s and $\xi_C$'s are taken as free parameters and determined
  by the $\chi^2$ fit in each individual category.
\end{enumerate}
Here we have taken the decay constants $f_\pi = 130.7$ MeV, $f_K = 159.8$ MeV
\cite{PDG}, $f_{K^*} = 210.4$ MeV and $f_\rho = 210.4$ MeV
\cite{vecdecayconst}. For Scheme 3 in the $DV$ sector, it turns out that this
scheme does not work well with the present available experimental data. We will
discuss this issue in Subsection \ref{sec:DV}.

Among all $B_{u,d}$ decays considered in this work, no Cabibbo-suppressed decay
involves the exchange diagram.  The only place to test this is the $B_s$
decays, of which we know very little at the moment.  We thus assume $\xi_E$'s
=1 when we predict the branching ratios of those decays.

The strong phases given in the following results are subject to a two-fold
ambiguity.  This is because only the cosines of the relative strong phases are
involved in the branching ratios.  Therefore, it is allowed to flip the signs
of all the phases simultaneously without changing the fitting quality and our
predictions.  In view of this, we will restrict the strong phase associated
with the color-suppressed amplitudes to the $[-180^\circ,0^\circ]$ range in our
analysis.

\subsection{$B \to D P$ decays}
\label{sec:DP}

In Table~\ref{tab:DPfit}, we see that $\chi^2_{\rm min}$ is greatly reduced by
the introduction of the SU(3) breaking factors $\xi_T$ and $\xi_C$.  The
smallness of $\chi^2_{\rm min}$ in Schemes~2 and 3 also shows the consistency
of input observables.

\begin{table}[h]
\caption{$B\to DP$ decays.  Theoretical parameters are extracted from global
  $\chi^2$ fits in different schemes explained in the text.  The amplitude
  sizes are given in units of $10^{-6}$.  Predictions of branching ratios are
  made with $\xi_E=1$ and given in units of $10^{-4}$ unless otherwise noted.}
\label{tab:DPfit}
\begin{center}
\begin{tabular}{l c c c} \hline \hline
      & Scheme 1  & Scheme 2 & Scheme 3 \\
\hline\hline 
$|T|$ & $16.26^{+0.61}_{-0.68} $ & $13.74 \pm 0.45$ 
& $13.71\pm 0.46 $ \\
$|C|$ & $6.77^{+0.20}_{-0.21} $ & $6.67 \pm 0.20 $
& $6.57 \pm 0.22 $\\
$|E|$ & $1.47^{+0.13}_{-0.15} $ & $1.48^{+0.13}_{-0.15} $
& $1.49^{+0.13}_{-0.15}$\\
$\delta_C$ (degrees) & $-69.0^{+9.2}_{-7.5}$ & $-47.0^{+9.5}_{-8.2}$
& $-48.7^{+9.8}_{-8.5} $ \\
$\delta_E$ (degrees) & $-146.2^{+13.9}_{-12.0}$ & $30.4^{+11.6}_{-11.8}$ &
$28.9^{+11.9}_{-12.1}$ \\
$\xi_T$  &1 (fixed) & $f_K/f_\pi$ (fixed) & $1.24 \pm 0.02 $ \\
$\xi_C$ &1 (fixed) & $f_K/f_\pi$ (fixed) & $1.33 \pm 0.02 $ \\
\hline
$\chi^2_{\rm min}$ & 45.28 & 3.53 & 1.41 \\
$\chi^2_{\rm min}/{\rm dof}$ & 11.32 & 0.88 & 0.71 \\
\hline\hline
$B^- \to D^0 \pi^-$ & $52.8 \pm 5.3 $ &  $ 48.6 \pm 3.7 $ & $ 47.5 \pm 3.8 $ \\
$\to D^0 K^-$ & $2.84 \pm 0.28 $ &  $ 3.91 \pm 0.30 $ & $ 4.08 \pm 0.34 $ \\
\hline
$\overline{B}^0 \to D^+ \pi^-$ & $ 29 \pm 3 $ &  $ 29 \pm 2 $ & $ 29 \pm 2 $ \\
$\to D^+ K^-$ & $ 1.8 \pm 0.1 $ &  $ 1.9 \pm 0.1 $ & $ 2.0 \pm 0.1 $ \\
$\to D^0 \pi^0$ & $ 2.76 \pm 0.37 $ &  $ 2.68 \pm 0.35  $ & $ 2.61 \pm 0.36 $ \\
$\to D^0 \eta$ & $ 2.2 \pm 0.3 $ &  $ 2.1 \pm 0.2 $ & $ 2.1 \pm 0.2 $ \\
$\to D^0 \eta'$ & $ 1.06 \pm 0.12 $ &  $ 1.03 \pm 0.11 $ & $ 1.00 \pm 0.12 $ \\
$\to D^0 \overline{K}^0$ & $ 0.31 \pm 0.02 $ &  $ 0.45 \pm 0.03 $ & $ 0.52 \pm 0.04 $ \\
$\to D_s^+ K^-$ & $ 0.27 \pm 0.05 $ &  $ 0.27 \pm 0.05  $ & $ 0.27 \pm 0.05 $ \\
\hline
$\overline{B}^0_s \to D^+ \pi^-$ (in units of $10^{-6}$) & $ 1.4 \pm 0.3 $ & $ 1.4 \pm 0.3
$ & $ 1.4 \pm 0.3 $ \\
$\to D^0 \pi^0$ (in units of $10^{-6}$) & $ 0.7 \pm 0.1 $ & $ 0.7 \pm 0.1 $ & $ 0.7 \pm 0.1 $ \\
$\to D^0 K^0$ & $ 5.4 \pm 0.3 $ & $ 5.3 \pm 0.3 $ & $ 5.1 \pm 0.3 $ \\
$\to D^0 \eta$  & $ 0.09 \pm 0.01 $ & $ 0.14 \pm 0.01 $ & $ 0.16 \pm 0.02 $ \\
$\to D^0 \eta'$  & $ 0.20 \pm 0.02  $ & $ 0.29 \pm 0.02 $ & $ 0.33 \pm 0.03 $ \\
$\to D_s^+ \pi^-$  & $ 31 \pm 2 $ & $ 22 \pm 1 $ & $ 22 \pm 1 $ \\
$\to D_s^+ K^-$  & $ 1.8 \pm 0.1 $ & $ 2.0 \pm 0.1 $ & $ 2.0 \pm 0.1 $ \\
\hline\hline
\end{tabular}
\end{center}
\end{table}

The values of $|T|$ and $|C|$ can be directly obtained from the $\overline{B^0}
\to D^+ K^-$ and $D^0 \overline{K^0}$ decays via the U-spin symmetry, i.e.,
exchange between $d$ quark and $s$ quark.  They are respectively $14.0 \pm 2.1$
and $7.17 \pm 0.46$ in units of $10^{-6}$ GeV. Here we take
$\xi_T=\xi_C=f_K/f_\pi$.  Likewise, $|E|$ is inferred from the $\overline{B^0}
\to D_s^+ K^-$ mode to be $(1.49 \pm 0.14) \times 10^{-6}$ GeV. These values
directly extracted from individual modes are consistent with those given in
Table~\ref{tab:DPfit} and in general have larger errors except for $|E|$. The
SU(3) breaking parameter $\xi_T$ can also be extracted from $\overline{B^0}\to
D^+K^-$ and $\overline{B_s^0}\to D_s^+\pi^-$. It leads to $\xi_T=0.96\pm0.22$.
This is smaller than the fitted value of $\xi_T$ in Table~\ref{tab:DPfit}.

According to our wave functions for $\eta$ and $\eta'$, the ratio ${\cal
  B}(D^0\eta) / {\cal B}(D^0\eta')$ is predicted to be 2, in comparison with
$1.58 \pm 0.33$ given by the current data.  From these decays, we determine
$|C+E| = (7.07 \pm 0.30) \times 10^{-6}$ GeV.  On the other hand, $|C-E| =
(6.42 \pm 0.30) \times 10^{-6}$ GeV is inferred from the $D^0 \pi^0$ mode.
Therefore, one can form the combination $|C|^2 + |E|^2 = (45.6 \pm 2.9) \times
10^{-12}$ GeV$^2$ from these three modes, consistent with $(53.6 \pm 6.6)
\times 10^{-12}$ GeV$^2$ that is derived from the $D^0 \overline{K^0}$ and
$D_s^+ K^-$ modes assuming $\xi_C=f_K/f_\pi$.

\begin{figure}[h]
\includegraphics[height=4.5cm]{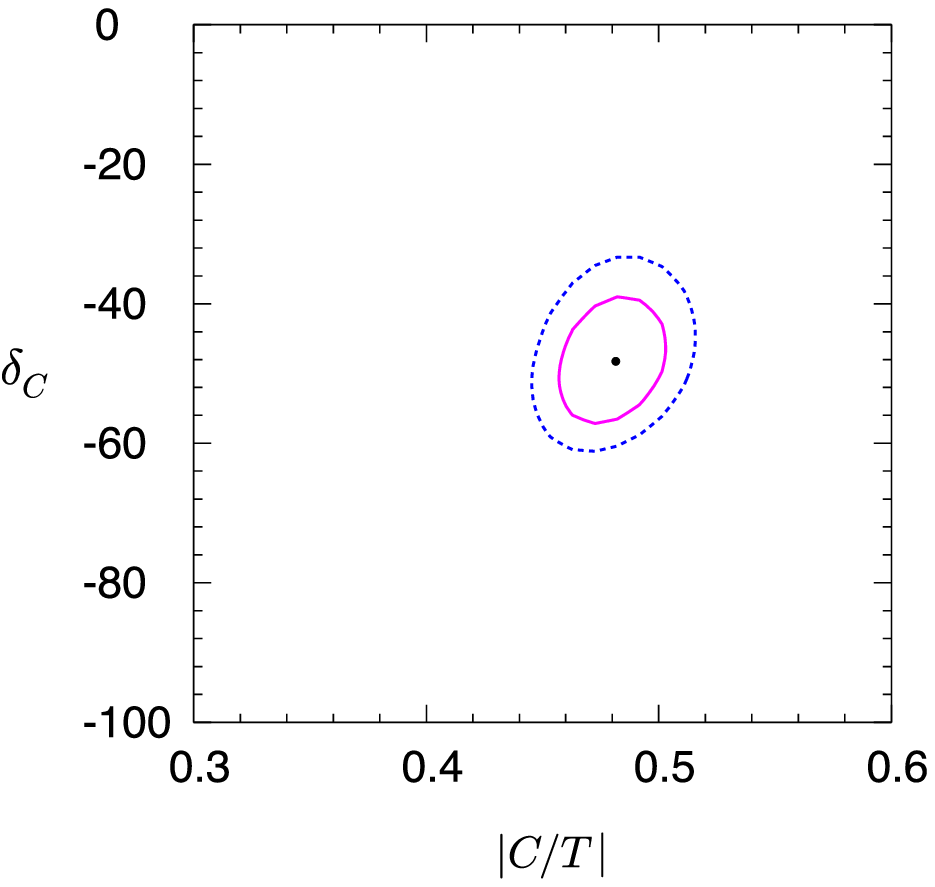} 
\hspace{0.3cm}
\includegraphics[height=4.5cm]{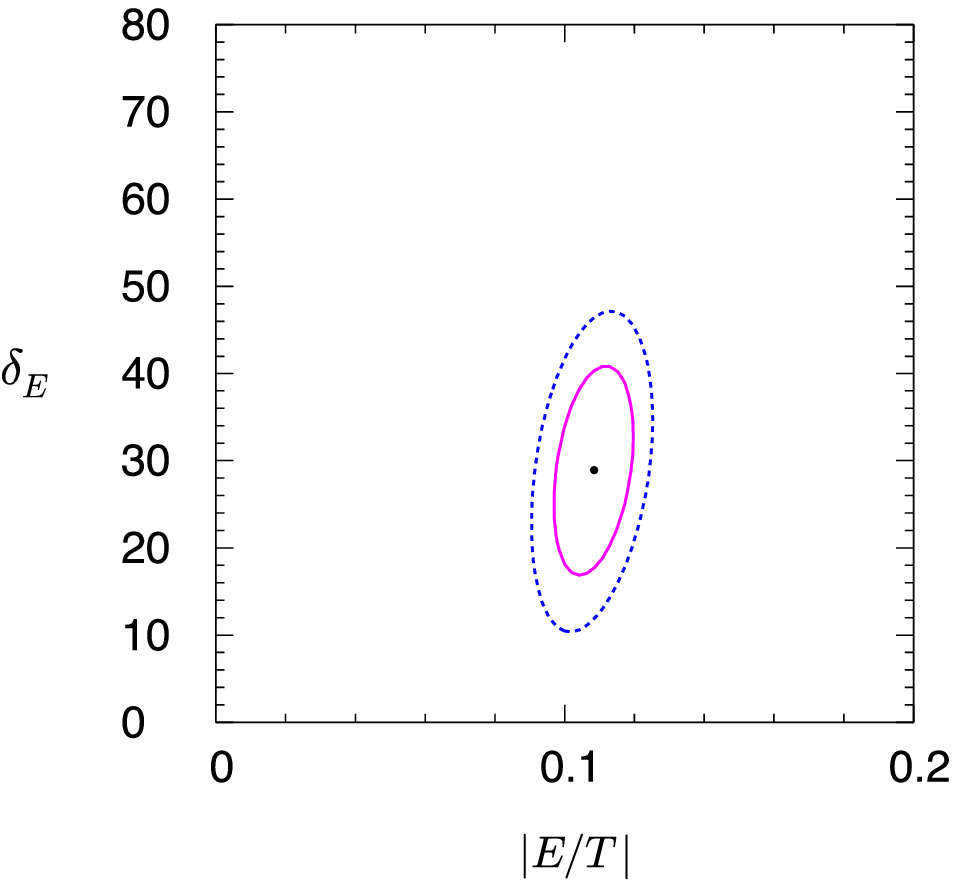}
\hspace{0.3cm}
\includegraphics[height=4.5cm]{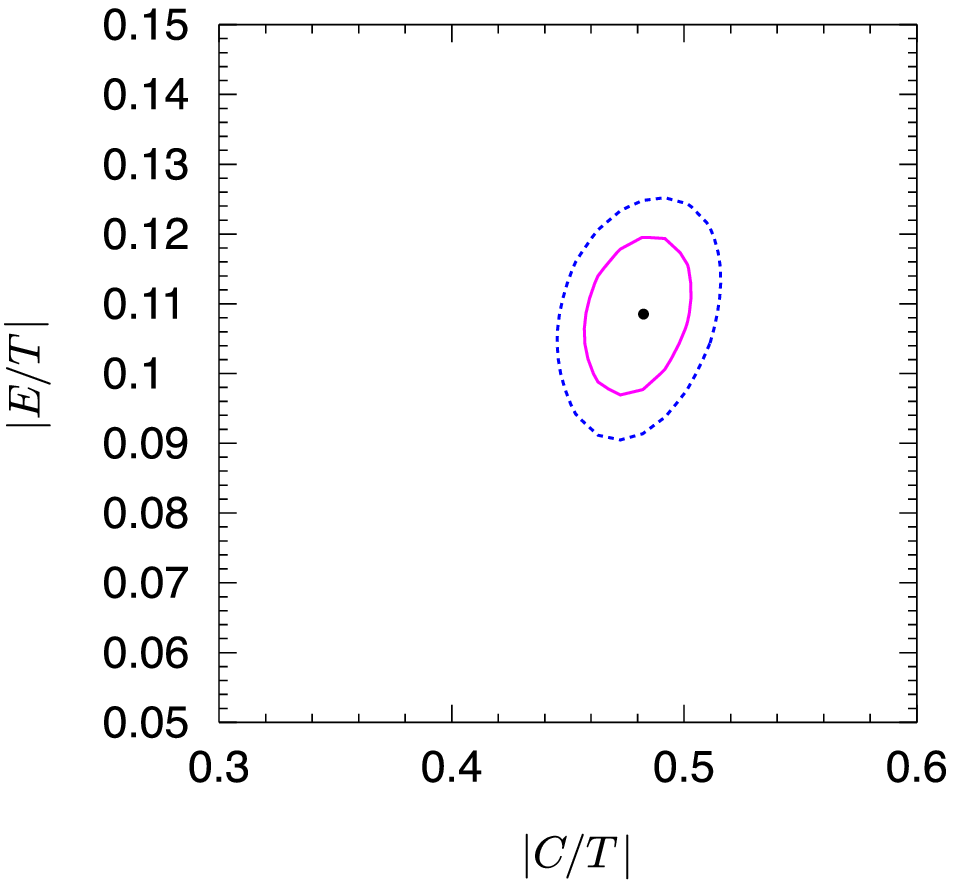} 
\caption{$\Delta\chi^2$=1 (pink, solid) and 2.30 (blue, dotted) contours on the
  $\delta_C$-$|C/T|$, $\delta_E$-$|E/T|$, and $|E/T|$-$|C/T|$ planes in Scheme
  3.}
\label{fig:DP2CE}
\end{figure}

In Fig.~\ref{fig:DP2CE}, we show the $\Delta \chi^2 = 1$ and $2.30$ contours on
the $\delta_C$-$|C/T|$, $\delta_E$-$|E/T|$, and $|E/T|$-$|C/T|$ planes in
Scheme 3, respectively, showing the correlations between each pair of
parameters.  The projections of the $\Delta \chi^2 = 1$ contours to individual
axes give the 68.3\% confidence level (CL) ranges of the corresponding
quantities.  In particular, we find that $|C/T| = 0.48\pm 0.02 $ and $|E/T| =
0.11\pm 0.01$.  Our result shows an enhancement in the color-suppressed
amplitude.  This can be explained by non-factorizable effects or final state
interactions.  The three flavor amplitude sizes fall into a hierarchy: $|T| >
|C| > |E|$, with $|E|$ being about one order of magnitude smaller than $|T|$.
This is the reason why the 1 $\sigma$ bounds on $\delta_E$ are relatively
loose.  Moreover, we observe non-trivial strong phases $\delta_C$ and
$\delta_E$.  These results are consistent with previous
studies~\cite{Chiang:2002tv,Kim:2004hx,Colangelo:2005hh}.

We note in passing that in our contour plots, the planes are scanned by
minimizing $\chi^2$, keeping all the other parameters free to vary. Therefore,
our results are different from those given in Ref.~\cite{Colangelo:2005hh}.  In
addition, their formalism corresponds to our Scheme 1.

Based upon our fit results, we give predictions for all the $B_{u,d,s}$ meson
decays in this category in the lower part of Table~\ref{tab:DPfit}.  For the
$B_s$ decays involving the exchange diagram, we take $\xi_E=1$. The predicated
branching ratios of those modes could be changed if we take into account SU(3)
breaking in $E$. Conversely, measurements of those modes can provide useful
information about the magnitude of the SU(3) breaking effect in the exchange
diagram.

$\overline{B_s}^0 \to D_s^+ \pi^-$ is a Cabibbo-favored decay involving the
tree amplitude.  Therefore, it has the largest decay rate among the channels in
this group.  Our preferred value for its branching ratio is $(22 \pm 1) \times
10^{-4}$. On the other hand, a recent measurement of this mode by CDF gives
$(38 \pm 3 \pm 13) \times 10^{-4}$~\cite{unknown:2006qw}.  The discrepancy is
1.2~$\sigma$. Further measurements of this and other $B_s$ decay modes with
better precision will help settling the question whether flavor SU(3) symmetry
can be reliably extended to the sector of $B_s$ meson decays or not.

From the naive factorization (NF) approximation, the SU(3) breaking parameters
are given by
\begin{eqnarray}
  \xi_T^{\rm NF} = \frac{f_K F_0^{BD}(m_K^2)}{f_\pi F_0^{BD}(m_\pi^2)} 
  \simeq 1.23~,
  &\quad&
  \xi_C^{\rm NF} = \frac{(m_B^2 - m_K^2) F_0^{BK}(m_D^2)}
  {(m_B^2 - m_\pi^2) F_0^{B\pi}(m_D^2)} \simeq 1.37 ~.
\end{eqnarray}
where the form factors are calculated using the covariant light-front
model~\cite{Cheng:2003sm}:
$F_0^{BD}(m_\pi^2)=0.67,~F_0^{BD}(m_K^2)=0.67,~F_0^{B\pi}(m_D^2)=0.28,
~F_0^{BK}(m_D^2)=0.38.$ These theoretical predictions are very close to our
fitted values: $\xi_T=1.24 \pm 0.02$ and $\xi_C=1.33 \pm 0.02$.

The ratio of the two effective Wilson coefficients $a_{1, 2}^{\rm eff}$ for
these decay processes can be extracted as
\begin{eqnarray}
  \left|\frac{a_2^{\rm eff}}{a_1^{\rm eff}}\right|_{DP} 
  &=& \left|\frac{C}{T}\right|\frac{(m_B^2 - m_D^2) f_\pi F_0^{BD}(m_\pi^2)}
  {(m_B^2 - m_\pi^2) f_D F_0^{B\pi}(m_D^2)}
  = 0.59 \pm 0.03~,
\end{eqnarray}
where $|C/T|=0.48\pm0.02$ as obtained from the $\chi^2$ analysis in Scheme 3,
and $f_D = 222.6$ MeV~\cite{PDG} is used.  In Ref.~\cite{Kim:2004hx},
$|a_2^{\rm eff}/a_1^{\rm eff}|_{DP}$ is found to be $0.54-0.70$ at the 1
$\sigma$ level using the data of $B^-\to D^0\pi^-$, $\overline{B}^0 \to D^+
\pi^-$ and $\overline{B}^0 \to D^0 \pi^0$ modes, which is consistent with our
result.  In the pQCD calculation~\cite{Keum:2003js}, it is found that
$|a_2^{\rm eff}/a_1^{\rm eff}|_{DP} =0.42-0.51$, and the relative phase between
$a_1^{\rm eff}$ and $a_2^{\rm eff}$ is estimated to be $-65.3^\circ<\arg(a^{\rm
  eff}_2/a^{\rm eff}_1)_{DP} <-61.5^\circ$ without the exchange diagram.

\subsection{$B \to D^* P$ decays}
\label{sec:DstP}

We see again in Table~\ref{tab:DstPfit} that $\chi^2_{\rm min}$ is
significantly lowered by the introduction of the SU(3) breaking factors
$\xi_{T_V} $ and $\xi_{C_P}$.

\begin{table}[h]
\caption{$B\to D^*P$ decays.  Theoretical parameters are extracted from global
  $\chi^2$ fits in different schemes explained in the text.  The amplitude
  sizes are given in units of $10^{-6}$.  Predictions of branching ratios are
  made with $\xi_E=1$ and given in units of $10^{-4}$ unless otherwise noted.}
\label{tab:DstPfit}
\begin{center}
\begin{tabular}{l c c c} \hline \hline
      & Scheme 1 & Scheme 2 & Scheme 3  \\
\hline\hline 
$|T_V|$ & $16.45^{+0.55}_{-0.61} $ & $14.85^{+0.60}_{-1.00} $ 
& $15.34^{+0.84}_{-1.70} $ \\
$|C_P|$ & $6.03^{+0.43}_{-0.46} $ & $6.21^{+0.39}_{-0.43} $
& $6.14^{+0.46}_{-0.50} $\\
$|E_P|$ & $1.37^{+0.18}_{-0.20} $ & $1.26^{+0.20}_{-0.23} $
& $1.29^{+0.19}_{-0.22}$\\
$\delta_{C_P}$ (degrees) & $-63.4^{+13.2}_{-10.8} $ & $-54.5^{+24.9}_{-12.0} $ 
& $-57.3^{+30.0}_{-12.6}$  \\
$\delta_{E_P}$ (degrees) & $-126.8^{+21.4}_{-19.3}$ & $-84.7^{+84.7}_{-27.8}$ &
$-100.9^{+100.9}_{-30.0}$ \\
$\xi_{T_V}$  & 1 (fixed) & $f_K/f_\pi$ (fixed) & $1.17^{+0.03}_{-0.02} $ \\
$\xi_{C_P}$ & 1 (fixed) & $f_K/f_\pi$ (fixed) & $1.20 \pm 0.05 $ \\
\hline
$\chi^2_{\rm min}$ & 12.00 & 1.39 & 0.72 \\
$\chi^2_{\rm min}/{\rm dof}$ & 3.00 & 0.35 & 0.36 \\
\hline\hline
$B^- \to D^{*0} \pi^-$ & $ 52 \pm 6 $ &  $ 49 \pm 8 $ & $ 50 \pm 10 $ \\
$\to D^0 K^-$ & $ 2.8 \pm 0.3 $ &  $ 3.9 \pm 0.6 $ & $ 3.7 \pm 0.8 $ \\
\hline
$\overline{B}^0 \to D^{*+} \pi^-$ & $ 30.3 \pm 2.8 $ &  $ 27.9 \pm 5.4 $ & $ 28.4 \pm 7.3 $ \\
$\to D^{*+} K^-$ & $1.80 \pm 0.13 $ &  $ 2.19 \pm 0.24  $ & $ 2.14 \pm 0.37 $ \\
$\to D^{*0} \pi^0$ & $ 1.9 \pm 0.5 $ &  $ 1.6 \pm 0.6 $ & $ 1.7 \pm 0.9 $ \\
$\to D^{*0} \eta$ & $ 1.9 \pm 0.4 $ &  $ 2.2 \pm 0.4 $ & $ 2.1 \pm 0.6 $ \\
$\to D^{*0} \eta'$ & $0.89 \pm 0.17 $ &  $ 1.05 \pm 0.21 $ & $ 1.00 \pm 0.29 $ \\
$\to D^{*0} \overline{K}^0$ & $ 0.24 \pm 0.04 $ &  $ 0.38 \pm 0.05 $ & $ 0.36 \pm 0.06 $ \\
$\to D_s^{*+} K^-$ & $ 0.23 \pm 0.06 $ &  $ 0.19 \pm 0.06 $ & $ 0.20 \pm 0.06 $ \\

\hline
$\overline{B}^0_s \to D^{*+} \pi^-$ (in units of $10^{-6}$) & $1.2 \pm 0.3 $ & $ 1.0 \pm 0.3
$ & $ 1.1 \pm 0.3 $ \\
$\to D^{*0} \pi^0$ (in units of $10^{-7}$) & $ 6.0 \pm 1.6 $ & $ 5.0 \pm 1.7 $ & $ 5.3 \pm 1.7 $ \\
$\to D^{*0} K^0$  & $ 4.2 \pm 0.6 $ & $ 4.5 \pm 0.6 $ & $ 4.4 \pm 0.7 $ \\
$\to D^{*0 }\eta$ & $ 0.06 \pm 0.02  $ & $ 0.09 \pm 0.02 $ & $ 0.09 \pm 0.04 $ \\
$\to D^{*0} \eta'$  & $ 0.16 \pm 0.03 $ & $ 0.27\pm 0.03 $ & $ 0.25 \pm 0.05 $ \\
$\to D_s^{*+} \pi^-$ & $ 31 \pm 2 $ & $ 25 \pm 3 $ & $ 27 \pm 4 $ \\
$\to D_s^{*+} K^-$  & $ 1.8 \pm 0.1 $ & $ 2.1 \pm 0.3 $ & $ 2.2 \pm 0.5 $ \\
\hline\hline
\end{tabular}
\end{center}
\end{table}

In this category, $|T_V| = (14.7 \pm 0.7) \times 10^{-6}$ GeV,
$|C_P|=(6.0\pm1.0)\times 10^{-6}$ GeV and $|E_P| =(1.29 \pm 0.21) \times
10^{-6}$ GeV can be directly extracted from the $D^{*+} K^-$, $D^{*0}
\overline{K^0}$ and $D_s^{*+} K^-$ modes respectively, taking
$\xi_{T_V}=\xi_{C_P}=f_K/f_\pi$.  Another way to constrain $|C_P|$ is to deduce
from the $D^{*0} (\pi,\eta,\eta')$ and $D_s^{*+} K^-$ modes.  Using this
method, we find $|C_P| = (6.2 \pm 0.5) \times 10^{-6}$ GeV.

\begin{figure}[h]
\includegraphics[height=4.5cm]{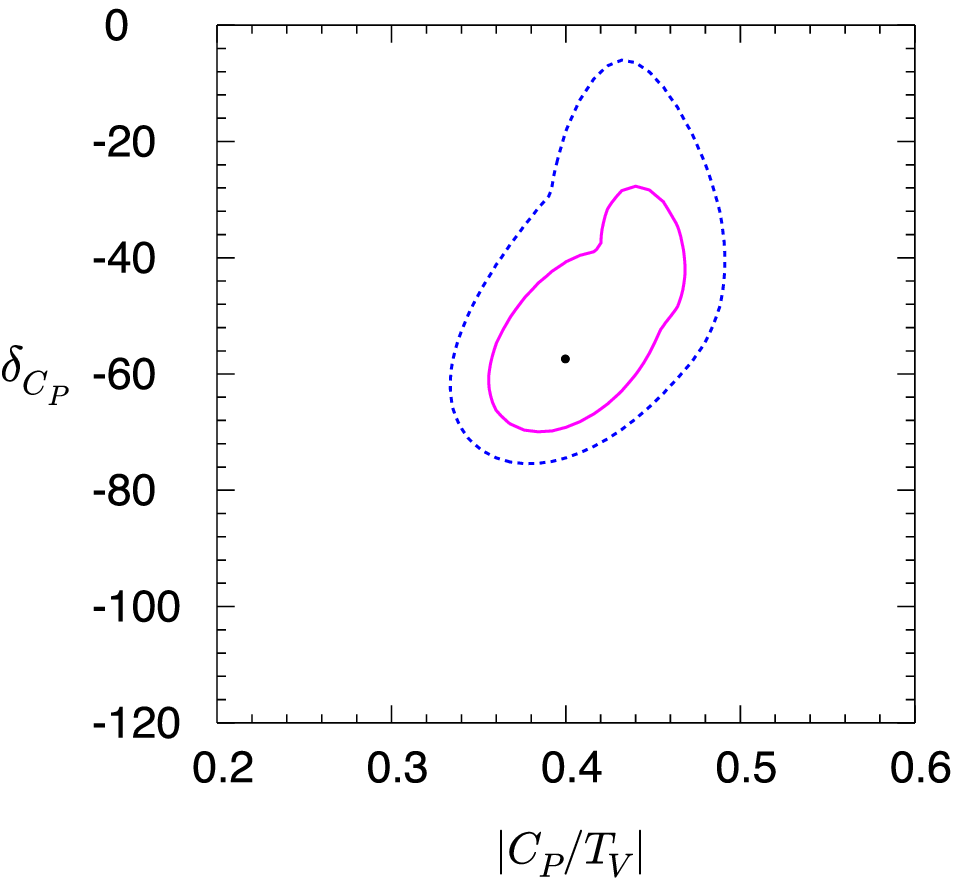} 
\hspace{0.3cm}
\includegraphics[height=4.5cm]{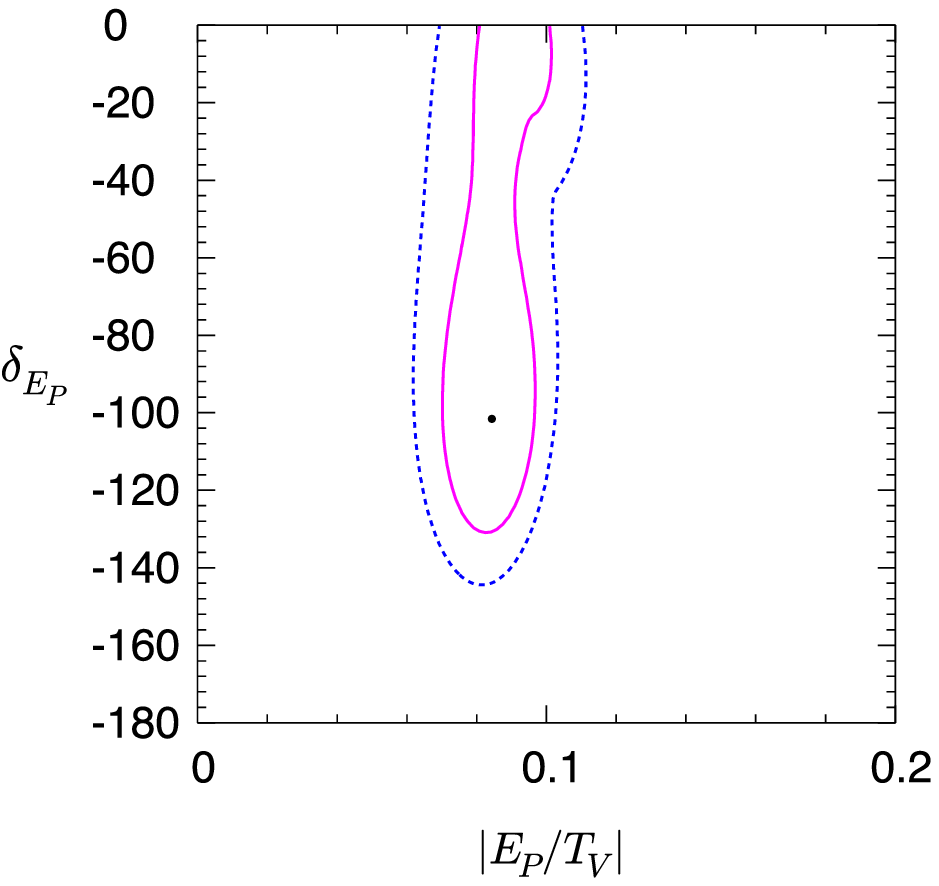} 
\hspace{0.3cm}
\includegraphics[height=4.5cm]{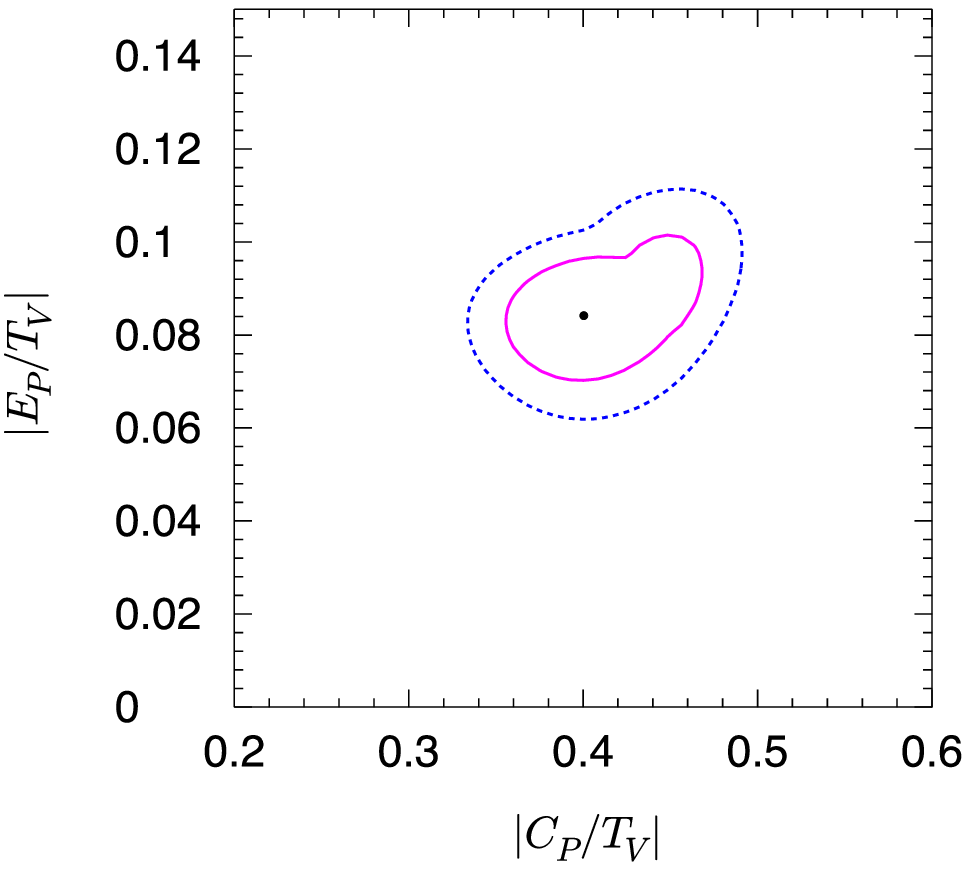}
\caption{$\Delta\chi^2$=1 (pink, solid) and 2.30 (blue, dotted) contours on the
  $\delta_{C_P}$-$|C_P/T_V|$, $\delta_{E_P}$-$|E_P/T_V|$, and
  $|E_P/T_V|$-$|C_P/T_V|$ planes in Scheme 3.}
\label{fig:DstP2CE}
\end{figure}

In Fig.~\ref{fig:DstP2CE}, we show the $\Delta \chi^2 = 1$ and $2.30$ contours
on the $\delta_{C_P}$-$|C_P/T_V|$, $\delta_{E_P}$-$|E_P/T_V|$, and
$|E_P/T_V|$-$|C_P/T_V|$ planes in Scheme 3, respectively.  We find that
$|C_P/T_V| = 0.40^{+0.07}_{-0.04}$ and $|E_P/T_V| = 0.08^{+0.02}_{-0.01}$. As
in the $DP$ category, there can be sizable non-factorizable contributions to
the color-suppressed amplitude or final state interactions.  The hierarchy among
$|T_V|$, $|C_P|$ and $|E_P|$ is also very similar to those in the $DP$ category
in Section~\ref{sec:DP}. However, the strong phase, $\delta_{E_P}$ can be zero
within the 68.3\% CL region.

Predictions for all the $B_{u,d,s}$ meson decays according to our fit results
are listed in the lower part of Table~\ref{tab:DstPfit}.  The most dominant
mode $D_s^{*+} \pi^-$ of $B_s$ decays is predicted to have a branching ratio of
$(27 \pm 4) \times 10^{-4}$, similar to that of the $D_s^+ \pi^-$ mode.

From the naive factorization approximation,
the SU(3) breaking parameters are given by
\begin{eqnarray}
  \xi_{T_V}^{\rm NF}=\frac{p^*_{D^*K} f_{K} A_0^{BD^*}(m_{K}^2)}
  {p^*_{D^*\pi} f_\pi A_0^{BD^*}(m_\pi^2)}  \simeq 1.22 ~,
  &\quad&
  \xi_{C_P}^{\rm NF}=\frac{p^*_{D^*K}F_1^{BK}(m_{D^*}^2)}
  {p^*_{D^*\pi}F_1^{B\pi}(m_{D^*}^2)}  \simeq 1.36,
\end{eqnarray}
where $A_0^{BD^*}(m_\pi^2)=0.64$, $A_0^{BD^*}(m_{K}^2)=0.65$,
$F_1^{B\pi}(m_{D^*}^2)=0.31$ and
$F_1^{BK}(m_{D^*}^2)=0.43$~\cite{Cheng:2003sm}.  Unlike the $DP$ sector, our
fitted SU(3) breaking factors are somewhat smaller than the naive factorization
expectations.

The ratio of the two effective Wilson coefficients can be extracted as
\begin{eqnarray}
  \left|\frac{a_2^{\rm eff}}{a_1^{\rm eff}}\right|_{D^*P}
  &=& \left|\frac{C_P}{T_V}\right|
        \frac{f_\pi A_0^{BD^*}(m_\pi^2)}{f_{D^*}F_1^{B\pi}(m_{D^*}^2)}
  = 0.42 \pm 0.04 ~,
\end{eqnarray}
where $|C_P/T_V|=0.40^{+0.07}_{-0.04}$ as obtained from the $\chi^2$ analysis
in Scheme 3, and $f_{D^*}$ = 256.0 MeV~\cite{PDG,Neubert:1997uc} is used.  In
the pQCD approach~\cite{Keum:2003js}, $|a_2^{\rm eff}/a_1^{\rm eff}|_{D^*P}$ is
found to be $0.47-0.55$ and their relative phase is estimated to be
$-64.8^\circ<\arg(a^{\rm eff}_2/a^{\rm eff}_1)_{D^*P} <-61.4^\circ$ without the
exchange diagram.  These are consistent with our results at the 1~$\sigma$
level.

\subsection{$B \to D V$ decays}
\label{sec:DV}

The decays in this category render a very different pattern from the previous
two in the $\chi^2$ fitting.  First, Scheme~1 and Scheme~2 yield the same
result.  This is because $f_{K^*}/f_\rho \simeq 1.00$.  Furthermore Scheme 3
does not work well, unlike the $DP$ and $D^*P$ sectors.  It is found that
$\chi_{\rm
  min}^2=0.045,~\xi_{T_P}=0.83,~\xi_{C_V}=4.58,~|T_P|=31.49,~|C_V|=1.75$ and
$|E_V|=6.64$ in units of $10^{-6}$ GeV, if we take $\xi_{T_P}$ and $\xi_{C_V}$
as free fitting parameters.  Theoretically, we do not expect $|C_V|<|E_V|$.
This unreasonable result is partly caused by the fact that $|E_V|$ is less
constrained by the experiment, $\overline{B}^0\to D_s^+K^{*-}$.  Therefore we
here adopt another prescription in which $\xi_{T_P}$ and $\xi_{C_V}$ are fixed
by the naive factorization calculation, i.e.,
\begin{eqnarray}
  \xi_{T_P}=\xi_{T_P}^{\rm NF} = \frac{p^*_{DK^*} f_{K^*} F_1^{BD}(m_{K^*}^2)}
  {p^*_{D\rho} f_\rho F_1^{BD}(m_\rho^2)} \simeq  1.00~,
  &~&
  \xi_{C_V}=\xi_{C_V}^{\rm NF} 
  = \frac{p^*_{DK^*}A_0^{BK^*}(m_D^2)}{p^*_{D\rho}A_0^{B\rho}(m_D^2)} 
  \simeq 1.09~,
\end{eqnarray}
where $F_1^{BD}(m_\rho^2)=0.69$, $F_1^{BD}(m_{K^*}^2)=0.69$,
$A_0^{B\rho}(m_D^2)=0.35$ and $A_0^{BK^*}(m_D^2)=0.38$~\cite{Cheng:2003sm}.

\begin{table}[h]
\caption{$B\to DV$ decays.  Theoretical parameters are extracted from global
  $\chi^2$ fits in different schemes explained in the text.  The amplitude
  sizes are given in units of $10^{-6}$.  Predictions of branching ratios are
  made with $\xi_E=1$ and given in units of $10^{-4}$ unless otherwise noted.}
\label{tab:DVfit}
\begin{center}
\begin{tabular}{l c c c} \hline \hline
      & Scheme 1 & Scheme 2 & Scheme 3  \\
\hline\hline 
$|T_P|$ & $25.60^{+1.56}_{-1.62} $ & $25.60^{+1.56}_{-1.62} $
& $25.87^{+1.61}_{-1.72} $ \\
$|C_V|$ & $7.07^{+0.29}_{-0.33} $ & $7.07^{+0.29}_{-0.33} $
& $6.95^{+0.29}_{-0.37}$\\
$|E_V|$ & $0.57^{+1.32}_{-0.43} $ & $0.57^{+1.32}_{-0.43} $
& $0.77^{+1.53}_{-0.66} $\\
$\delta_{C_V}$ (degrees) & $-75.1^{+19.1}_{-15.8} $ & $-75.1^{+19.1}_{-15.8} $
&$ -79.2^{+18.0}_{-14.9} $ \\
$\delta_{E_V}$ (degrees) & $143.4^{+36.6}_{-108.8}$ & $143.4^{+36.6}_{-108.8}$ &
$158.6^{+21.4}_{-128.5}$ \\
$\xi_{T_P}$ & 1 (fixed)  & $f_{K^*}/f_\rho$ (fixed) & $ \xi_{T_P}^{\rm NF}$ (fixed) \\
$\xi_{C_V}$ & 1 (fixed) & $f_{K^*}/f_\rho$ (fixed) & $ \xi_{C_V}^{\rm NF}$ (fixed) \\
\hline
$\chi^2_{\rm min}$ & 5.91 & 5.91 & 4.18 \\
$\chi^2_{\rm min}/{\rm dof}$ & 2.96 & 2.96 & 2.09 \\
\hline\hline
$B^- \to D^0 \rho^-$ & $105 \pm 18 $ &  $ 105 \pm 18 $ & $ 103 \pm 18 $ \\
$\to D^0 K^{*-}$ & $ 5.7 \pm 1.0 $ &  $ 5.7 \pm 1.0 $ & $ 5.6 \pm 1.0 $ \\
\hline
$\overline{B}^0 \to D^+ \rho^-$ & $ 78 \pm 11 $ &  $ 78 \pm 11 $ & $ 78 \pm 12 $ \\
$\to D^+ K^{*-}$ & $ 4.3 \pm 0.5 $ &  $ 4.3 \pm 0.5 $ & $ 4.4 \pm 0.6 $ \\
$\to D^0 \rho^0$ & $ 3.5 \pm 0.8 $ &  $ 3.5 \pm 0.8 $ & $ 3.4 \pm 1.0 $ \\
$\to D^0 \omega$ & $ 2.7 \pm 0.7 $ &  $ 2.7 \pm 0.7 $ & $ 2.7 \pm 0.9 $ \\
$\to D^0 \overline{K}^{*0}$ & $ 0.33 \pm 0.03 $ &  $ 0.33 \pm 0.03 $ & $ 0.38 \pm 0.04 $ \\
$\to D_s^+ K^{*-}$ & $ 0.04 \pm 0.12 $ &  $ 0.04 \pm 0.12 $ & $ 0.07 \pm 0.20 $ \\
\hline
$\overline{B}^0_s \to D^+ \rho^-$ (in units of $10^{-7}$) & $ 2.1 \pm 6.3 $ & $ 2.1 \pm 6.3 
$ & $ 3.8 \pm 10.7 $ \\
$\to D^+ K^{*-}$  & $ 4.2 \pm 0.6 $ & $ 4.2 \pm 0.6 $ & $ 4.2 \pm 0.6 $ \\
$\to D^0 \rho^0$ (in units of $10^{-6}$) & $ 1.9 \pm 5.8 $ & $ 1.9 \pm 5.8  $ & $ 3.5 \pm 9.8 $ \\
$\to D^0 K^{*0}$  & $ 5.8 \pm 0.5 $ & $ 5.8 \pm 0.5 $ & $ 5.6 \pm 0.5 $ \\
$\to D^0 \omega$ (in units of $10^{-7}$) & $ 1.0 \pm 3.2 $ & $ 1.0 \pm 3.2  $ & $ 1.9 \pm 5.4 $ \\
$\to D^0 \phi$  & $ 0.31 \pm 0.03 $ & $ 0.31 \pm 0.03 $ & $ 0.35 \pm 0.03 $ \\
$\to D_s^+ \rho^-$ & $ 75 \pm 9 $ & $ 75 \pm 9 $ & $ 77 \pm 10 $ \\
$\to D_s^+ K^{*-}$  & $ 4.1 \pm 0.6 $ & $ 4.1 \pm 0.6 $ & $ 4.2 \pm 0.6 $ \\
\hline\hline
\end{tabular}
\end{center}
\end{table}

$|T_P| = (26.1 \pm 2.0) \times 10^{-6}$ GeV can be extracted from the $D^+
K^{*-}$ mode using the U-spin symmetry and taking $\xi_{T_P}=f_{K^*}/f_\rho$.
This is slightly larger than our fit result in Scheme 2.  Directly from the
$\overline{B^0} \to D_s^+ K^{*-}$ mode, we have only a poor upper bound of $8.2
\times 10^{-6}$ GeV on $|E_V|$.

The observable ${\cal B}(D^0 \rho^-)$ has the largest contribution to the total
$\chi^2_{\rm min}$.  From Table \ref{tab:DV}, we observe that the area of the
triangle formed from the $B^-\to D^0\rho^-$, $\overline{B}^0 \to D^+ \rho^-$
and $\overline{B}^0 \to D^0 \rho^0$ decays is very small, while that of the
triangle formed from the $B^-\to D^0K^{*-}$, $\overline{B}^0 \to D^+ K^{*-}$
and $\overline{B}^0 \to D^0 \overline{K}^{*0}$ modes is not.  This is the
reason why the global $\chi^2$ fits in the $DV$ sector are not as satisfactory
as those in the $DP$ and $D^*P$ sectors.

In Ref.~\cite{Chiang:2002tv}, we noted that $|C_V|$ extracted from
$D^0\overline{K}^{*0}$ was inconsistent with $\sqrt{|C_V|^2+|E_V|^2}$ extracted
from a combination of the $D^0\rho^0$ and $D^0\omega$ modes.  Currently, the
former is $(8.01 \pm 0.60) \times 10^{-6}$ GeV if we take
$\xi_{C_V}=f_{K^*}/f_\rho$, and the latter is $(6.86 \pm 0.34) \times 10^{-6}$
GeV.  There is still a discrepancy at the 1.7$\;\sigma$ level, or this
discrepancy implies that the SU(3) breaking factor $\xi_{C_V}$ should be
greater than about $1.17$.  A determination of ${\cal B}(D_s^+ K^{*-})$ and
better measurements of related modes will be very useful in providing further
insights into this problem.

\begin{figure}[h]
\includegraphics[height=4.5cm]{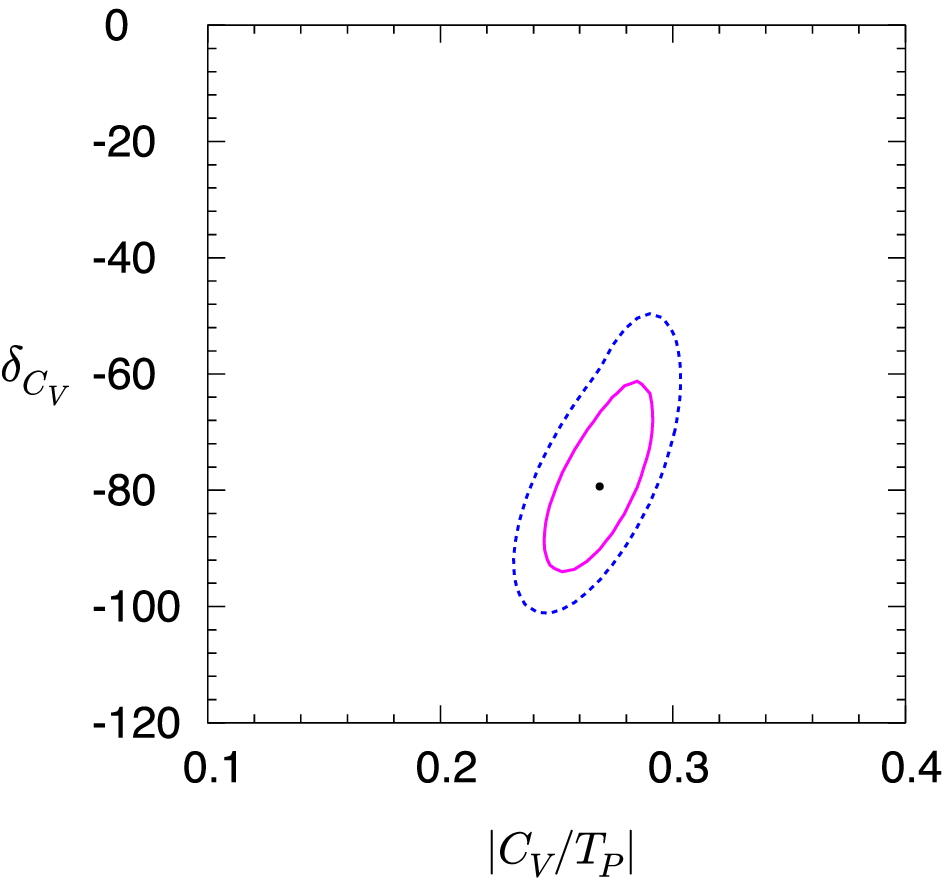} 
\hspace{0.3cm}
\includegraphics[height=4.5cm]{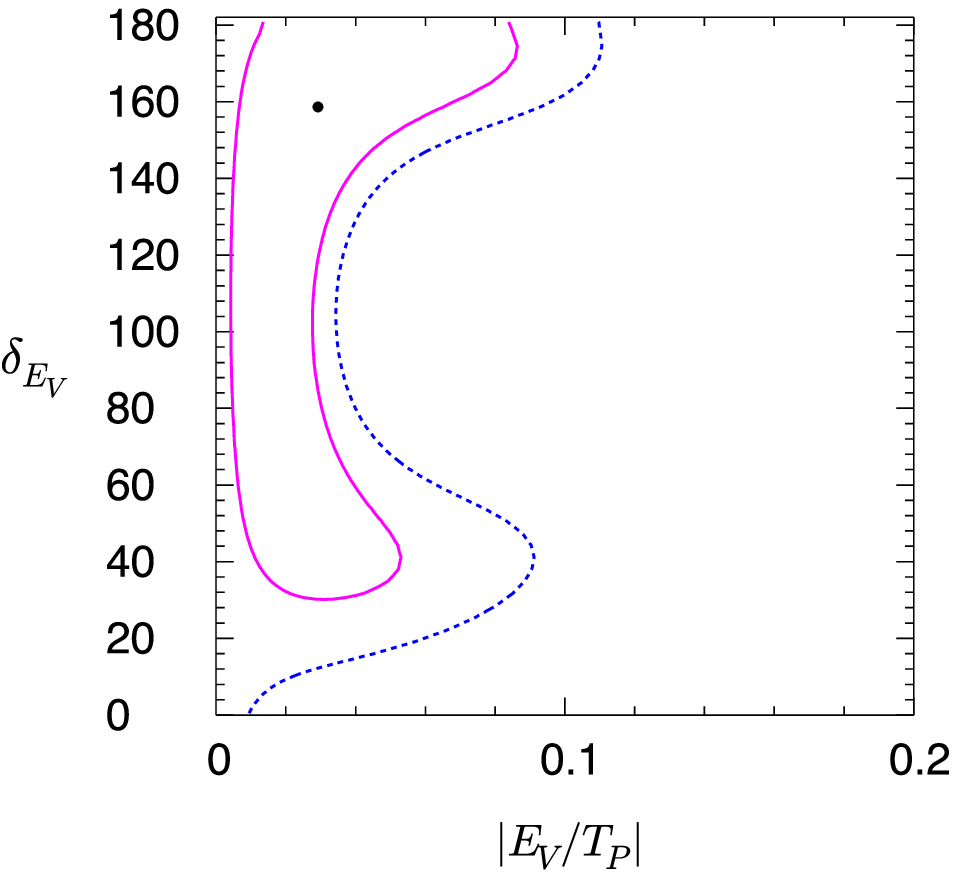} 
\hspace{0.3cm}
\includegraphics[height=4.5cm]{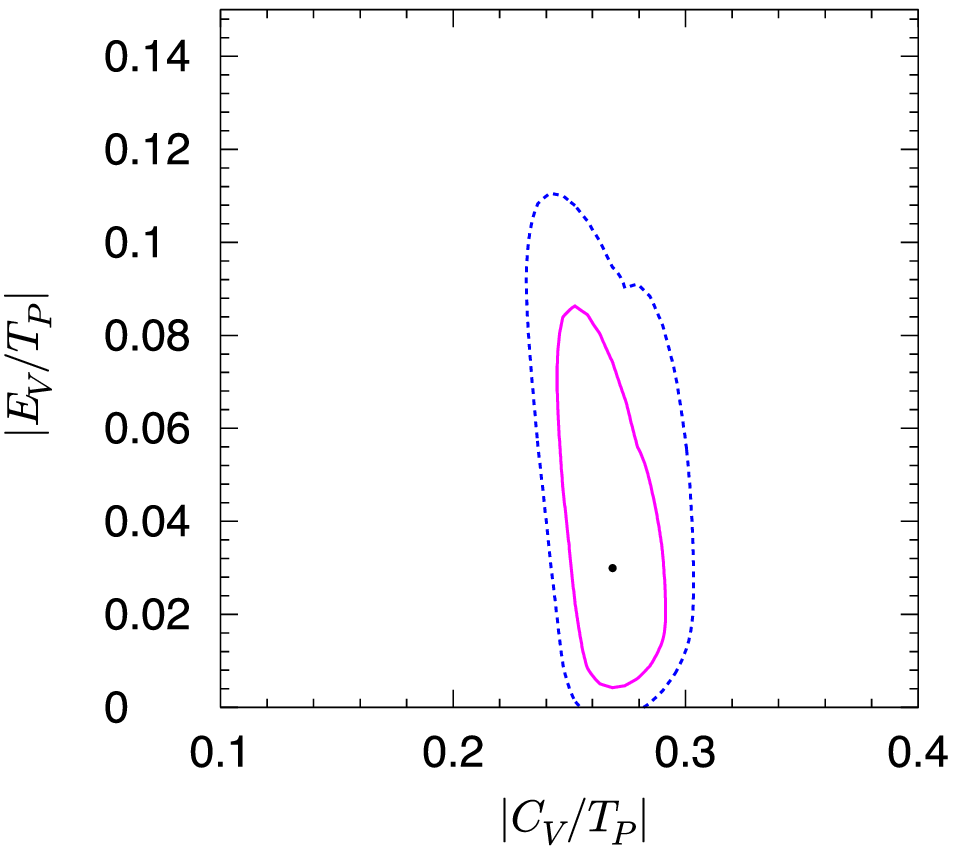} 
\caption{$\Delta\chi^2$=1 (pink, solid) and 2.30 (blue, dotted) contours on the
  $\delta_{C_V}$-$|C_V/T_P|$, $\delta_{E_V}$-$|E_V/T_P|$, and
  $|E_V/T_P|$-$|C_V/T_P|$ planes in Scheme 3.}
\label{fig:DV2CE}
\end{figure}

In Fig.~\ref{fig:DV2CE}, we show the $\Delta \chi^2 = 1$ and $2.30$ contours on
the $\delta_{C_V}$-$|C_V/T_P|$, $\delta_{E_V}$-$|E_V/T_P|$, and
$|E_V/T_P|$-$|C_V/T_P|$ planes in Scheme 3, respectively.  We find that
$|C_V/T_P| =0.27 \pm 0.02$ and $|E_V/T_P| =0.03^{+0.06}_{-0.03}$.  We see that
the magnitude of $T_P$ is larger than $T$ and $T_V$, resulting in a more
hierarchical structure among $|T_P|$, $|C_V|$ and $|E_V|$.  Another result of
the large $T_P$ is reflected in the bigger branching ratio prediction for the
most dominant $\overline{B_s}^0 \to D_s^+ \rho^-$ mode. As in the $D^*P$
sector, the central value of $\delta_{E_V}$ is non-zero, but is still
consistent with zero within the 68.3\% CL region.

The ratio of the two effective Wilson coefficients can be extracted as
\begin{eqnarray}
  \left|\frac{a_2^{\rm eff}}{a_1^{\rm eff}}\right|_{DV}
  &=& \left|\frac{C_V}{T_P}\right|\frac{f_\rho F_1^{BD}(m_\rho^2)}
  {f_D A_0^{B\rho}(m_D^2)}
  = 0.50\pm 0.04  ~,
\end{eqnarray}
where $|C_V/T_P| =0.27 \pm 0.02$ as obtained from the $\chi^2$ analysis in
Scheme 3. In Ref.~\cite{Kim:2004hx}, it is estimated that $|a_2^{\rm
  eff}/a_1^{\rm eff}|_{DV}=0.24-0.42$ at the 1~$\sigma$ level using the data of
the $B^-\to D^0\rho^-$, $\overline{B}^0 \to D^+ \rho^-$ and $\overline{B}^0 \to
D^0 \rho^0$ modes.

\begin{figure}[h]
\begin{center}
\includegraphics[height=3cm]{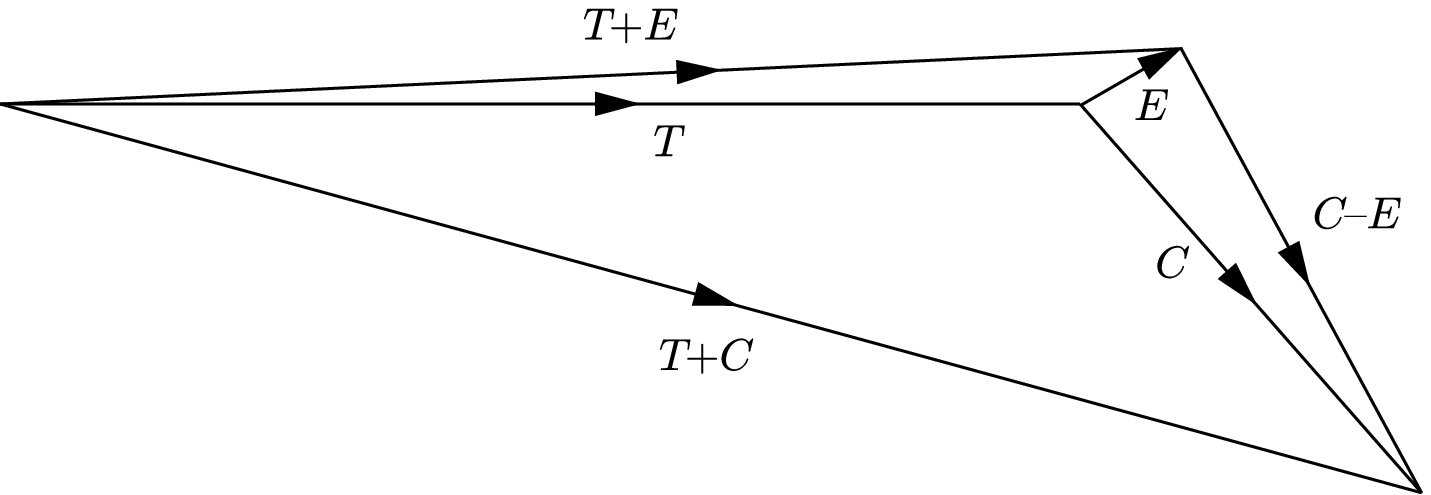} \\
(a) \\[0.5cm]
\includegraphics[height=3cm]{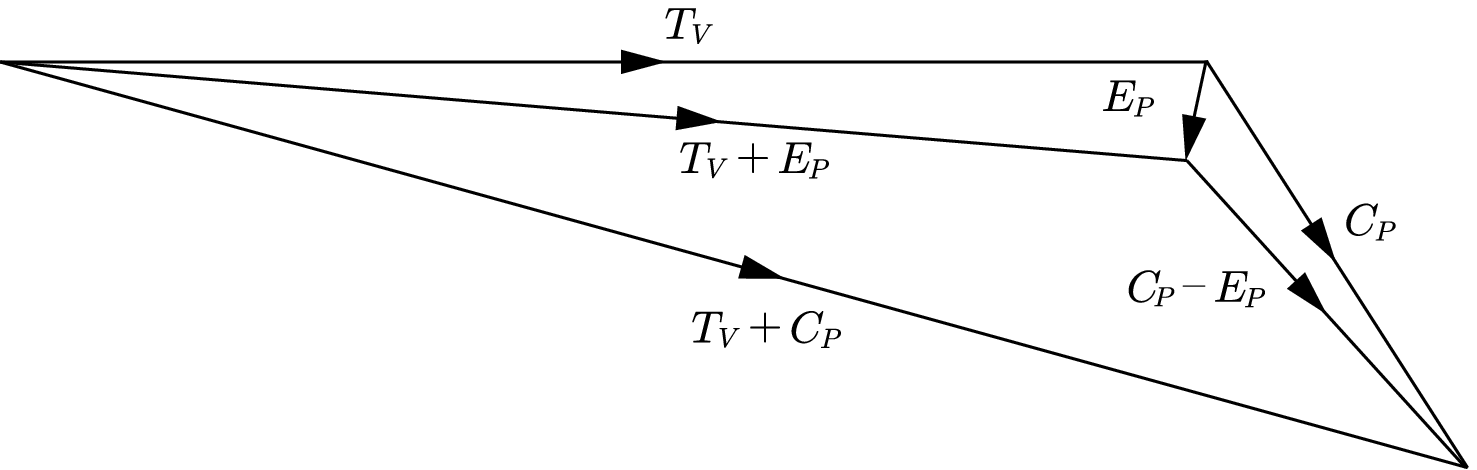} \\
(b) \\[0.5cm]
\includegraphics[height=3cm]{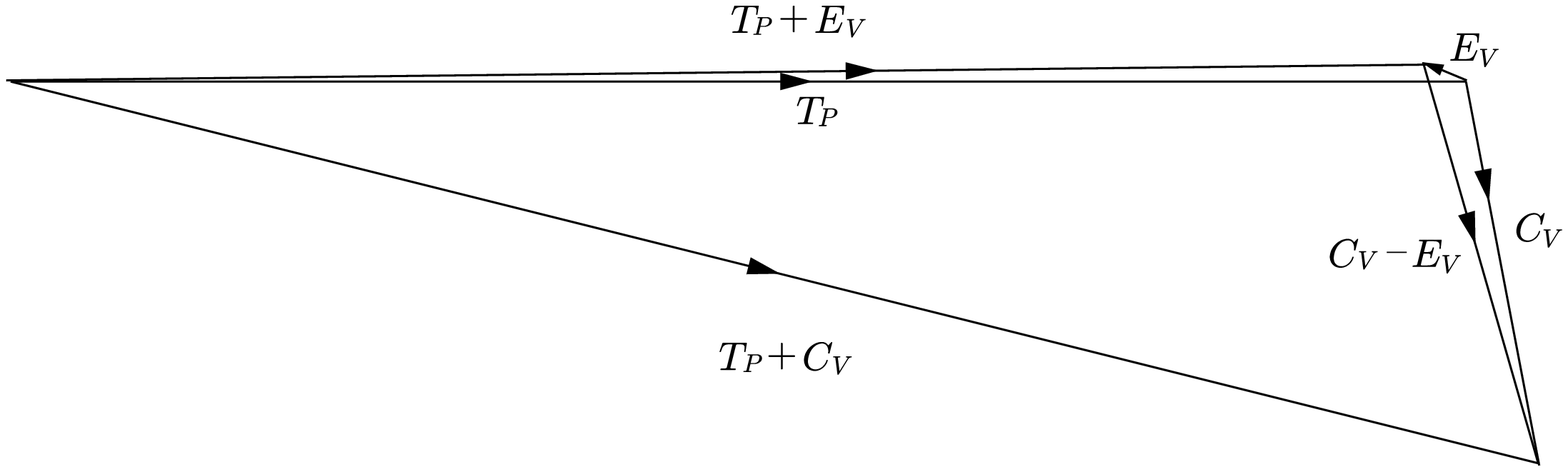} \\
(c) \\[0.5cm]
\caption{Amplitude diagrams of (a): $DP$ decays; (b): $D^*P$ decays; and (c):
  $DV$ decays.
\label{fig:amps}}
\end{center}
\end{figure}

Finally, we summarize our findings in Fig.~\ref{fig:amps}.  These diagrams are
constructed by taking the central values of the fitted parameters in each
category using Scheme 3.  They illustrate the sizes and relative phases among
the tree, color-suppressed, and exchange amplitudes.

\section{Conclusions \label{sec:summary}}

We have used the $\chi^2$ fit approach to re-analyze the two-body charmed $B$
meson decays in the flavor SU(3)-symmetric formalism, taking into account
different symmetry breaking schemes as well.  In the $DP$ and $D^* P$ decays,
there are significant improvement in the $\chi^2$ minimum between Scheme~1 and
Scheme~2, but not much between Scheme~2 and Scheme~3.  This shows that the
major SU(3) breaking effect can be accounted for by the decay constant ratio
$f_K/f_\pi$, as demanded for example by naive factorization.  The same feature,
however, is not observed in the $DV$ sector, where the corresponding decay
constant ratio is approximately one.

In our analysis, the fit results are generally consistent with those extracted
from individual modes.  We have found that the color-suppressed amplitudes are
enhanced in the $DP$ and $D^*P$ sectors, but not in the $DV$ sector.  This
strongly suggests that non-factorizable effects or final-state rescattering
effects cannot be neglected in the former two sectors.

In the $DV$ sector, it is observed that the Cabibbo-suppressed $D^0
\overline{K}^{*0}$ yields a $|C_V|$ that exceeds the bound $\sqrt{|C_V|^2 +
  |E_V|^2}$ given by a combination of the $D^0 \rho^0$ and $D^0 \omega$
branching ratios at $1.7~\sigma$ level, or $\xi_{C_V}$ should be greater than
about $1.17$. We urge the measurement of ${\cal B}(\overline{B}^0\to
D_s^+K^{*-})$ for a direct determination of the exchange amplitude, which may
provide a possible solution to this problem.

Finally we note that the exchange diagrams are at least an order of magnitude
smaller than the dominant tree topologies in these decays.  Consequently, it is
difficult to determine their phases, particularly in the $D^*P$ and $DV$
sectors, unless data precision can be significantly improved in the future.

\begin{acknowledgments}
  C.-W.~C. would like to thank the hospitality of the National Center for
  Theoretical Sciences in Taiwan and the Institute of Theoretical Physics at
  Univ.\ of Oregon during his visit where part of this work was initiated and
  carried out.  This research was supported in part by the National Science
  Council of Taiwan, R.O.C.\ under Grant No.\ NSC 95-2112-M-008-008.
\end{acknowledgments}


\end{document}